\documentclass[11pt,prd,superscriptaddress,nofootinbib]{revtex4}
\usepackage[english]{babel}
\usepackage{graphicx}
\usepackage{mathtext}
\usepackage{indentfirst}
\usepackage{epsfig,amsmath,amsfonts}
\usepackage{amsmath,amssymb,pgfplots}
\usepackage{braket}

\newcommand{\beq}[1]{\begin{eqnarray}\label{#1}}
	\newcommand\eeq {\end{eqnarray}}
\newcommand\bqa {\begin{eqnarray}}
	\newcommand\eqa {\end{eqnarray}}

\newcommand{\bear}{\begin{array}}
	\newcommand{\enar}{\end{array}}

\begin{document}
	\begin{titlepage}
	\hfill ITEP-TH-01/16 \hspace*{3.4mm}
	\vspace{2.5cm}
	\begin{center}
		
		\centerline{\Large \bf A way to distinguish very compact stellar objects from black holes}
		
		\vskip 1.5cm {Emil T.\ Akhmedov$^{1,2}$, Daniil A. Kalinov$^{1}$ and
			Fedor K.\ Popov$^{1,2,3}$}
		\\
		{\vskip 0.5cm
			$^{1}$National Research University Higher School of Economics, Russian Federation, \\
          International Laboratory of Representation Theory and Mathematical Physics,\\
			\vskip 0.5cm
			$^{2}$B. Cheremushkinskaya, 25,
			Institute for Theoretical and Experimental Physics,\\
			117218, Moscow, Russian Federation}
		\vskip 0.5cm
		$^{3}$Institutskii per, 9, Moscow Institute of Physics and Technology,\\
		141700, Dolgoprudny, Russian Federation
		
		\vskip 0.5cm
	\end{center}
	
	\vskip 0.35cm
	
	\begin{center}
		\today
	\end{center}
	
	\noindent
	
	\vskip 1.2cm
	\centerline{\bf Abstract}
	We propose a way to distinguish compact stellar object, whose size is very close to its Schwarzschild
radius, from the collapsing stars. Namely, we show that {\it massive} fields in the vicinity of a very compact stellar object have discrete energy levels. (These levels are different from the standard non-relativistic ones present in Coulomb type of potentials and from the quasinormal modes.) At the same time we show that there are no such discrete levels for massive fields in the vicinity of a collapsing star.
	\end{titlepage}

\section{Introduction}

There are several methods that allow one to distinguish compact stellar objects (such as e.g. neutron stars) from black holes (see e.g. \cite{Abramowicz:2011xu}, \cite{Postnov:2014tza}).
Standard neutron star models predict that their size should be substantially grater than the corresponding
Schwarzschild radius, $r_g$ (in this mote we restrict our attention to the non--rotating case). At least it should be greater than the last stable circular orbit, $r = 3r_g/2$. However,
suppose that there is such an exotic state of matter\footnote{For the options consider e.g. \cite{Mottola:2010gp}, \cite{Mottola:2011ud}, \cite{Mazur:2015kia} (see also \cite{Nicolini:2005vd}--\cite{Kawai:2013mda} and \cite{Virbhadra:1999nm}--\cite{Virbhadra:2008ws}).} which allows the existence of a compact stellar object, whose radius, $R$, is very close to the corresponding Schwarzschild scale,  $3r_g/2 > R \sim r_g$. Such an object is very hard to distinguish from the black hole by the standard methods due to the enormous infrared shift from its surface. In this note we propose a method that may allow one to show such a distinction.

	In particular, in \cite{Akhmedov:2015xwa} the existence of discrete energy levels for massive scalar particles in the vicinity of a compact massive thin shell at rest was predicted. In this paper we estimate their number and positions and extend their consideration to the case of interior solution \cite{Misner:1974qy} and to the collapse. These modes are different from the standard discrete levels present in nonrelativistic limit in the Coulomb/Newtonian like potentials. Also they are not the same as the quasinormal modes \cite{Kokkotas:1999bd}.

The presence of the discrete levels in the vicinity of a very compact stellar object can be easily predicted. In fact, very close to the Schwarzschild radius every field behaves as massless. Then, if the mass of the field is not zero, its effective potential in the gravitational background under consideration has a well in the vicinity of the star surface (see fig. (\ref{fig1})). The well is present, if the field has the regular boundary conditions either at the center of the star or on its surface. If, however, there is a collapse instead of the static star, the boundary conditions are changed such that one does not fetch stationary states. Rather one encounters running modes, which have continuous spectrum.
This is the difference one can use to distinguish very compact stellar objects from the black hole collapse.

 For the discussion that is very similar in spirit to our
paper, but different in details consider \cite{Chirenti:2007mk}. For the discussion of the bosonic stars and of the similar states (to those that we consider here) on the bosonic star backgrounds please see \cite{Brito:2015yfh}--\cite{Seidel:1993zk}. For the discussion of the similar states on the rotating black hole backgrounds and their physical implications please see \cite{Brito:2013wya}--\cite{Hod:2012zza}.

	\section{Thin-Shell at rest}

For illustrative reasons we start with the consideration of the scalar field theory in the background of a massive thin shell. The metric inside the shell is flat. At the same time outside it the metric has the Schwarzchild's form (see \cite{Israel:1966rt},\cite{Israel:1967zz} for the original work and, e.g., \cite{Poisson} for the review):
	\bqa\label{metric}
	ds^2 = \left\{\begin{matrix}
		ds_-^2 = dt_-^2 - dr^2 - r^2d\Omega^2, \quad & r  \leq R(t) & \\[2mm]
		ds_+^2 = \left(1 - \frac{r_g}{r}\right) dt^2 - \frac{dr^2}{1 - \frac{r_g}{r}} -  r^2d\Omega^2, \, & r \geq R(t) &
	\end{matrix}\right., \quad d\Omega^2 = d\theta^2 + \cos^2\theta d\varphi^2.
	\eqa
	Here $R(t)$ is the radial coordinate of the shell and $t(t_-)$ is the time coordinate outside (inside) the shell.
	
	For simplicity we consider the theory of a free real massive scalar field on such a background:
	\bqa
	 S = \int d^4 x \sqrt{|g|}\left[\frac{1}{2} \left(\partial_\mu\phi\right)^2  - \frac{1}{2} m^2 \phi^2\right].\label{scifield}
	\eqa
	Decomposing $ \phi = \sum\limits_{l,m}\phi_l(r,t) Y_{l,m}(\theta,\phi)$ with the real spherical functions $Y_{l,m}(\theta,\phi)$, we obtain for $\phi_l$ the following equations:
	\bqa\label{eoms}
	\begin{cases}
		\left[\partial_{t_-}^2 - \partial_r^2 + m^2 + \frac{l(l+1)}{r^2}\right] \, \left(r \phi_{l}\right) = 0, & \quad r \leq R(t) \\[2mm]
		\left[\partial_{t}^2 - \partial_{r^*}^2 + \left(1 - \frac{r_g}{r}\right)\left(m ^2 + \frac{l(l+1)}{r^2} + \frac{r_g}{r^3}\right)\right] \, \left(r \phi_{l}\right) = 0, & \quad r \geq R(t),
	\end{cases}
	\eqa
	where $r^* = r + r_g \log\left(\frac{r}{r_g} - 1\right)$ and the boundary conditions on the shell are as follows:
	\begin{align}\label{bound}
	\phi_{l}\Bigl[R(t) -  0 \Bigr] &= \phi_{l}\Bigl[R(t) +  0\Bigr], \notag\\[4mm]
	\left[\left(\frac{\partial t}{\partial t_-}\right)\, \left|\frac{dR}{dt}\right|\, \partial_{t} \phi_{l}   - \left(\frac{\partial t_-}{\partial t} \right) \partial_r \phi_{l}\right]_{r=R(t) - 0} &=
	\left[\frac{\partial_{t} \phi_{l}}{1 - \frac{r_g}{r}} \left|\frac{dR}{dt}\right|  - \left(1 - \frac{r_g}{r}\right) \, \partial_r \phi_{l}\right]_{r=R(t) + 0}.
	\end{align}
	They follow from the least action principle for (\ref{scifield}) in the metric (\ref{metric}) \cite{Akhmedov:2015xwa}. The second condition is the continuity of the normal derivative of the field $\phi$ across the shell. In this section we assume that $R(t)$ is time independent and is equal to $R_0$ due to the presence of some force. We consider it to be very small $|R_0 - r_g| \ll r_g$, but still greater than $r_g$. Such a situation models very compact stationary stellar object.

First, continuity of the metric, $ds^2_-  = ds^2_+$, and the stationarity of the shell, $dr=0$, provide from (\ref{metric}) the relation between inside and outside clock rates \cite{Akhmedov:2015xwa}: $t_- = \sqrt{1 - \frac{r_g}{r}} \,\,t$. Second, if one introduces the following coordinate,

\bqa\label{newcoordinate}
	\hat{r} = \left\{
	\begin{matrix}
	r \left(1 - \frac{r_g}{R_0}\right)^{-\frac12}  , r \leq R_0\\
	r + r_g \log\left(\frac{r}{r_g} - 1 \right) - R_0 - r_g \log\left(\frac{R_0}{r_g} - 1 \right) + R_0 \left(1 - \frac{r_g}{R_0}\right)^{-\frac12}, r \geq R_0,
	\end{matrix}
	\right.
	\eqa
	equations of motion and boundary conditions can be represented in a more compact form for $\psi_l = r \phi_l$:
	\begin{gather}\label{eqonpsi}
	\left[\partial_t^2 - \partial_{\hat{r}^2} + V_l(\hat{r})\right]\psi_l = 0.
	\end{gather}
	Where the potential is
	\begin{gather}
	V_l\left[\hat{r}(r)\right] = \left\{\begin{matrix}
		\left(1 - \frac{r_g}{R_0}\right)\left(m^2 + \frac{l(l+1)}{r^2}\right), r < R_0\\
		\left(1 - \frac{r_g}{r}\right)\left(m^2 + \frac{l(l+1)}{r^2} + \frac{r_g}{r^3}\right), r > R_0,
	\end{matrix}\right.
	\end{gather}
	and the boundary conditions are as follows
	\begin{gather}\label{boundarycond}
	\psi_l\left(\hat{R}_0 - 0\right) = \psi_l \left(\hat{R}_0+0\right),\quad{\rm and}\quad\partial_{\hat{r}}\psi_l\left(\hat{R}_0 - 0\right) = \partial_{\hat{r}} \psi_l \left(\hat{R}_0+0\right).
	\end{gather}
	Note that $\hat{r} \to +\infty$, as $r \to +\infty$; $\hat{r} = \frac{R_0}{\sqrt{1 - \frac{r_g}{R_0}}} \equiv \hat{R}_0$, when $r = R_0$; and $\hat{r} \to 0$, as $r\to 0$.

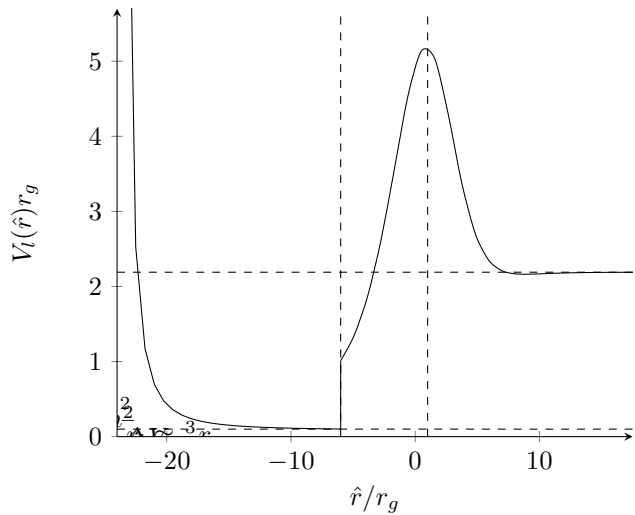
\begin{figure}[h!]
		\centering
	\begin{tikzpicture}
	\newcommand{\xrgmin}{-6}
		\newcommand\rgnum{2}
		\newcommand\massnum{1.3}
		\newcommand{\lnum}{10}
	\begin{axis}[
	axis lines=left,
	ylabel = $V_l(\hat{r}) r_g$,
	xlabel = $\hat{r}/r_g$,
	ymin=0,
	ymax=5.7,
	xmin=-24,
	xmax=17.5
	]
	\addplot[black,domain=-24:\xrgmin]{(1.69+\lnum*(\lnum+1)/(x+24)^2)*(0.05)};
	\addplot[black,smooth,domain=\xrgmin:20]{0.5+(1-1/(1 + e^(-1  + x/\rgnum)))*(1.69+\lnum*(\lnum+1)/(\rgnum + \rgnum*e^(-1+x/\rgnum))^2 + \rgnum/(\rgnum + \rgnum*e^(-1+x/\rgnum))^3};
	\addplot[black,dashed,domain=-24:17.5]{2.19} node[above,pos=0.33] {$m^2$};
	\addplot[black,dashed,domain=-24:17.5]{0.1} node[above,pos=0.33] {$m_-^2$};
	\addplot[black,dashed] coordinates{(-6,0) (-6,5.7)} node [left,pos=0.7]{$r=R_0$};
	\addplot[black] coordinates{(-6,0.1) (-6,1.01)} node [right,pos=0.5]{$\Delta V$};
	\addplot[black,dashed] coordinates{(1.0,0) (1.0,5.7)} node[right,pos=0.95]{$r_l \approx \frac{3}{2} r_g$};
	\end{axis}
	\end{tikzpicture}
		\caption{Qualitative plot of the potential $V(\hat{r})$. Note that this plot is valid for $l \gg 1$. We discuss the case of $l\sim 1$ below.}
		\label{fig1}
\end{figure}

We are looking for the solutions of \eqref{eqonpsi} of the form $\psi_l(\hat{r},t) = \bar{\psi}_{\omega,l} e^{- i \omega t}$. Then the single particle problem on the background in question reduces to:
	\begin{gather}\label{eigenvalue}
	\left[-\partial_{\hat{r}}^2 + V_l(\hat{r})\right] \psi_{\omega,l} = \omega^2 \psi_{\omega,l}.
	\end{gather}
For the beginning let us discuss the properties of the potential in question. From the fig. (\ref{fig1}) one can see that $V_l(\hat{r}) \to m^2$, as $ \hat{r} \to +\infty$ and $V_l(\hat{r}) \to \infty$, as $ \hat{r} \to 0$. The minimum of the potential is equal to $V_{\rm min} = V\left(\frac{R_0}{\sqrt{1 - \frac{r_g}{R_0}}} - 0\right) =  \left(m^2 + \frac{l(l+1)}{R_0^2}\right)\left(1 - \frac{r_g}{R_0}\right)$, at the same time the local maximum is $V_{\rm max} = V_l\left[\hat{r}(r_l)\right] \approx \frac{4 l^2}{27 r_g^2}$ and is reached at $r = r_l \approx \frac{3}{2} r_g$. Also at $r = R_0$ there is a jump in the potential of the magnitude $\Delta V_l = \left(1 - \frac{r_g}{R_0}\right) \frac{r_g}{R_0^3}$, which is very small for $|R_0 - r_g| \ll r_g$. This jump is present due to the change of the behaviour of the metric accros the shell. From the fig. (\ref{fig1}) it should be apparent that there are discrete levels in such a potential, if $m > 0$.
	
Let us investigate properties of the harmonics in (\ref{eigenvalue}). We have to demand their regularity at $\hat{r} = 0$ and $\hat{r} =+ \infty$. Henceforth, for $\omega^2 > m^2$ we have a continuous spectrum of such functions that are regular at $\hat{r} = 0$. They are discussed in \cite{Akhmedov:2015xwa} and are not of our interests in the present paper. At the same time for $m^2 > \omega^2 > m_-^2$, where $m^2_- = m^2 \left(1 - \frac{r_g}{R_0}\right)$, the spectrum is discrete and contains finitely many energy levels, as we will see in a moment.

To find these states we look for approximate solutions of the equation \eqref{eigenvalue} in the regions $0 \le r \le R_0$ and $ R_0 \le r$ and glue them across $r = R_0$ using boundary conditions \eqref{boundarycond}. In the region $R_0 \le r \lesssim \frac{3}{2} r_g$ the equation (\ref{eigenvalue}) can be approximated in the following way
\begin{gather}
\left[-\partial_{\hat{r}}^2 + e^{\hat{r}/r_g} k^2\right] \psi_{\omega,l} \approx \omega^2 \psi_{\omega,l},\quad{\rm where}\quad k^2 = e^{a}\left[m^2 + \frac{l(l+1) + 1}{r_g^2}\right].
\end{gather}
Here $a$ is a constant that determines the difference between $\hat{r}$ and $r^*$ and can be found from \eqref{newcoordinate}. Solution of this equation can be represented as a linear combination of Hankel functions of an imaginary index and argument:
\begin{gather}\label{linearsol}
\psi_{\omega,l}(\hat{r}) = C_1 H^{(1)}_{2i \omega r_g}\left(2 i k r_g e^{\frac{\hat{r}}{2 r_g}} \right) + C_2 H^{(2)}_{2i \omega r_g}\left(2 i k r_g e^{\frac{\hat{r}}{2 r_g}} \right).
\end{gather}
For the discrete levels the normalization conditions imply that $\psi_{\omega,l}(\hat{r})$ has to be finite as $\hat{r} \to + \infty$. Bessel functions for large values of their argument behave in the following way $H^{(1)/(2)}_{2i\omega r_g}(x) \sim e^{\pm i x}$. One can conclude that $C_2$ has to vanish, because in our case $x \propto  i e^{\frac{\hat{r}}{2 r_g}}$. In the quasiclassical approximation in the vicinity of the shell harmonic functions (\ref{linearsol}) with $C_2 = 0$ can be approximated as follows
\begin{gather}\label{firstapproximation}
\psi_{\omega,l}(\hat{r}) \approx A \sin\left[\sqrt{\omega^2 - m_-^2} \hat{r} + \varphi(\omega,r_g)\right],
\end{gather}
where $A,\varphi$ are independent of $\hat{r}$.  The constant $\varphi(\omega,r_g)$ is determined from the position of the turning point, that is close to $r \approx \frac{3}{2} r_g$. The constant $A$ follows from the normalization and gluing conditions. Also we have neglected the aforementioned small jump $\Delta  V$ of the potential at $r = R_0$.

In the region $ 0 \le r \le R_0$ the equation (\ref{eigenvalue}) can be approximated as:
\begin{gather}
\left[-\partial_{\hat{r}}^2 + \left(1 - \frac{r_g}{R_0}\right)\left(m^2 + \frac{l(l+1)}{r^2}\right)\right] \approx \omega^2 \psi_{l,\omega}(\hat{r}).
\end{gather}
Solution of this equation also is a linear combination of Bessel and Neumann functions,  $J_{l+\frac{1}{2}}(x)$ and $Y_{l+\frac{1}{2}}(x)$.  The condition of regularity at $\hat{r} = 0$ defines the  harmonics inside the shell up to a total factor:
\begin{gather}
\psi_{l,\omega}(\hat{r}) \approx B  \left[\left(\omega^2 - m^2\right)\hat{r}^2\right]^{\frac14} \, J_{l+\frac{1}{2}}\left(\sqrt{\omega^2 - m_-^2} \hat{r}\right),
\end{gather}
where $B$ is some constant following from the normalization condition \cite{Akhmedov:2015xwa}.
In the vicinity of the shell this function can be approximated as
\begin{gather}\label{secondapproximation}
\psi_{l,\omega}(\hat{r}) \approx B \sin\left(\sqrt{\omega^2 - m_-^2} \hat{r} -  \frac{\pi l}{2}\right).
\end{gather}
Gluing \eqref{firstapproximation} and \eqref{secondapproximation} at the shell, we get the condition $B = A$ and equality of the phases. The latter equality states just that there is an integer number of wavelengthes between turning points:
\bqa
\sqrt{\omega^2 - m_-^2} L \approx \pi n + \frac{\pi}{2},
\eqa
here $L \approx \hat{r}\left(\frac{3}{2}r_g\right)$ is the distance between the origin and the turning point.

The last equation defines the energy spectrum. For the large values of $n$ it can be approximated as

\bqa\label{energystates}
\omega_n^2 \approx m_-^2 + \frac{\pi^2 n^2}{L^2}.
\eqa
One can estimate the total number of discrete levels as $N \approx \frac{m L}{\pi}$ because the discrete spectrum does exist  only for $\omega_n < m$.

It is probably worth stressing at this point that these discrete states are different from the usual ones that appear in the non-relativistic approximation, i.e. in the Newtonian/Columb-like potential $U(r) \propto - \frac{1}{r}$ (for the case of large angular momentum $l \gg m r_g$). In fact,  in the limit $r \gg r_g$ the potential under consideration looks like
\begin{gather}
V(r) \approx m^2 - \frac{m^2 r_g}{r} + \frac{l(l+1)}{r^2}.
\end{gather}
We substitute this potential into the equation \eqref{eigenvalue}, where we neglect the difference between $\hat{r}$ and $r$, when $r \gg r_g$. Also we expand $\omega = m + \epsilon$ and assume that $\epsilon \ll m$. As the result we obtain the Schroedinger equation
\begin{gather}
\left[-\frac{\partial_{r}^2}{2m} - \frac{r_g\, m}{2\,r} + \frac{l(l+1)}{2\, r^2 \, m}\right]\psi_{\omega,l} = \epsilon \psi_{\omega,l}.
\end{gather}
The energy spectrum of this problem is well-known $\epsilon_n = - \frac{m^3 r_g^2}{ 8n^2}, n = l + n_r + 1 > l$, where $n_r$  is a radial quantum number \cite{LL}. If $l$ is sufficiently large, then our approximation of $\epsilon \ll m$ is valid and we can treat it in the non-relativistic way. These states are localized around $r_n \approx \frac{n^2}{m^2 r_g}\sim \frac{l^2}{m^2 r_g} \gg r_g$ (note the very small well right after the local maximum on the fig. (\ref{fig1})), while our states \eqref{energystates} are localized in the very vicinity of the shell: at $3r_g/2 > r \sim R \geq r_g$.

\begin{figure}[h!]
		\centering
	\begin{tikzpicture}
	\newcommand{\xrgmin}{-6}
		\newcommand\rgnum{2}
		\newcommand\massnum{1.3}
		\newcommand{\lnums}{4}
        \newcommand{\lnumss}{3}
        \newcommand{\lnumsss}{5}
	\begin{axis}[
	axis lines=left,
	ylabel = $V_l(\hat{r}) r_g$,
	xlabel = $\hat{r}/r_g$,
	ymin=0,
	ymax=5.7,
	xmin=-24,
	xmax=17.5
	]
	\addplot[black,domain=-24:\xrgmin]{(1.69+\lnums*(\lnums+1)/(x+24)^2)*(0.05)};
\addplot[black,domain=-24:\xrgmin]{(1.69+\lnumss*(\lnumss+1)/(x+24)^2)*(0.05)};
\addplot[black,domain=-24:\xrgmin]{(1.69+\lnumsss*(\lnumsss+1)/(x+24)^2)*(0.05)};
	\addplot[black,smooth,domain=\xrgmin:20]{(1-1/(1 + e^(-1  + x/\rgnum)))*(1.69+\lnums*(\lnums+1)/(\rgnum + \rgnum*e^(-1+x/\rgnum))^2 + \rgnum/(\rgnum + \rgnum*e^(-1+x/\rgnum))^3};
	\addplot[black,smooth,domain=\xrgmin:20]{(1-1/(1 + e^(-1  + x/\rgnum)))*(1.69+\lnumss*(\lnumss+1)/(\rgnum + \rgnum*e^(-1+x/\rgnum))^2 + \rgnum/(\rgnum + \rgnum*e^(-1+x/\rgnum))^3};
	\addplot[black,smooth,domain=\xrgmin:20]{(1-1/(1 + e^(-1  + x/\rgnum)))*(1.69+\lnumsss*(\lnumsss+1)/(\rgnum + \rgnum*e^(-1+x/\rgnum))^2 + \rgnum/(\rgnum + \rgnum*e^(-1+x/\rgnum))^3};
	\addplot[black,dashed,domain=-24:17.5]{1.69} node[above,pos=0.33] {$m^2$};
	\addplot[black,dashed,domain=-24:17.5]{0.1} node[above,pos=0.33] {$m_-^2$};
	\addplot[black,dashed] coordinates{(1.0,0) (1.0,5.7)} node[left,pos=0.9]{$r_l \approx \frac{3}{2} r_g$};
	\end{axis}
	\end{tikzpicture}
		\caption{Qualitative plot of potential $V(\hat{r})$ for several different values of $l \sim 1$.}
		\label{fig2}
\end{figure}
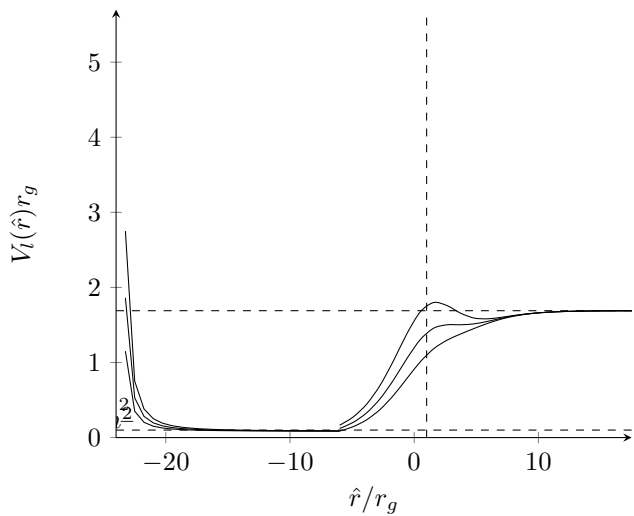

Note that for small values of $l$, when the nonrelativistic approximation is not applicable anymore, the second well merges with the first one, as it can be seen from the fig. (\ref{fig2}).

\section{The case of the interior solution}

The result of the previous section can be straightforwardly generalized to the case of interior solution \cite{Misner:1974qy}. For this solution the metric is read as follows:
\begin{gather}
ds^2 = e^{2 g} \tilde{f} dt^2 - \tilde{f}^{-1} dr^2 - r^2 d\Omega^2.
\end{gather}
Here  $\tilde{f}(r) =1 - \frac{2 m(r)}{r}$ and $e^{2 g(r)}$ together with $m(r)$ follow from the equations \cite{Misner:1974qy}:
\begin{gather}
\partial_r m = 4 \pi r^2 T^t_t \label{OSm} \\
\partial_r g = 4\pi r^2 \tilde f^{-1}(T^t_t - T^r_r) \notag
\end{gather}
Where $T^{\mu}_\nu = {\rm Diag}\left(\rho,-p,-p,-p\right)$.
We assume that the star's matter content obeys the equation of state of the form: $p = k \rho, 0 \le k \le 1$. Furthermore, the pressure and density are constant for $r\leq R_0$ and they are zero for $r>R_0$, i.e. we are dealing with the incompressible fluid matter content of the star. We restrict our attention to this type
of matter just for simplicity reasons. All our considerations below can be straightforwardly extended for the other types of matter.

For the matter under consideration equations \eqref{OSm} can be easily integrated to give
\begin{equation}
 \tilde{f} = \left\{\begin{matrix}
 1 -\frac{r_g r^2}{R_0^3}, r < R_0\\
 1 - \frac{r_g}{r}, r > R_0
 \end{matrix}\right. ,\quad {\rm and}\quad  e^{2 g} = \left(\frac{\tilde f}{1 - \frac{r_g}{R_0}}\right)^{-\frac{3}{2}\left(k+1\right)}, \label{solOS}
\end{equation}
where $r_g = \frac{8}{3} \pi G \rho R_0^3$. Finally, we assume that $\left|R_0-r_g\right|\ll r_g$ as in the previous section. Note that with the star matter content under consideration, the stability conditions demand that $R_0 > 9 r_g/8$ (see e.g. the discussion in \cite{Misner:1974qy}). However, other types of exotic matter allow to relax the latter condition.

As in the first section we consider the theory of the real massive scalar field \eqref{scifield}  on a such background. Also we  decompose the field in real spherical harmonics, $ \phi = \sum_{l,m} \phi_l(r,t) Y_{l,m}(\theta,\phi)$:
\begin{gather}\label{OSsol}
 S = \sum_{l,m} \frac{1}{2}\int^\infty_0 r^2 dr \int dt\left[\frac{1}{\tilde{f} e^{2g}} \left(\partial_t \phi_l\right)^2 - \tilde{f} \left(\partial_r \phi_l\right)^2 - \left(m^2 + \frac{l(l+1)}{r^2}\right) \phi_l^2\right].
 \end{gather}
 Varying \eqref{OSsol} we obtain:
\begin{gather}\label{OSqm}
\left[e^{-2g}\partial_t^2 - \partial^2_{r^*} + \tilde{f} \left(m^2 + \frac{l(l+1)}{r^2} + \frac{\partial_r \tilde{f}}{r}\right)\right] \psi_l  = 0,
\end{gather}
where we have introduced $\psi_l=r \phi_l$ and $r^* = \int^r_0 \frac{dr}{\tilde{f}(r)}$.

Again we are looking for the solutions of the form $\psi_l = e^{-i \omega t }\psi_{l,\omega}$ and the problem reduces to:
\bqa
\left[ - \partial^2_{r^*} + \tilde{f} \left(m^2 + \frac{l(l+1)}{r^2} + \frac{\partial_r \tilde{f}}{r}\right)\right] \psi_{l,\omega}  = e^{-2 g} \omega^2 \psi_{l,\omega}. \label{orthOS}
\eqa
This problem is similar to the one of the previous section, but now the potential inside the star has changed and depends on $\omega$ itself. For $\omega > m$ we also obtain the continuous spectrum, while for $\omega < m$ there is a possibility of existence of discrete energy levels.

First, note that under the above conditions we have that $e^{-2g}\ll 1$, when $r<R_0$. Hence, to estimate the energy levels we can neglect the right-hand side of the equation \eqref{orthOS} to obtain
\bqa
\left[ - \partial^2_{r^*} + \tilde{f} \left(m^2 + \frac{l(l+1)}{r^2} + \frac{\partial_r \tilde{f}}{r}\right)\right] \psi_l(r^*)  =0
\eqa
Therefore in the range $0 < r < R_0$ the harmonic functions have an $\omega$-independent approximate form, which we denote as $\psi_l(r^*)$.

Second, outside the star $e^{-2 g} = 1$ and the equation reduces to \eqref{eqonpsi} for $r > R_0$. Again in the vicinity of the star surface we can use the approximation \eqref{firstapproximation}, $\psi_{\omega,l}(r^*) \approx A \cos\left(\sqrt{\omega^2 - m_-^2} r^* + \phi'\right)$, for the behavior of the harmonics. Sewing logarithmic derivatives of $\psi_l(r^*)$ and $\psi_{\omega,l}(r^*)$ at $r^* = R^*_0$ we get the following conditions on the energy spectrum:
\begin{gather*}
\partial_{r^*} \log \psi_{\omega,l}(R^*_0 + 0) =\partial_{r^*} \log \psi_{l}(R^*_0 - 0) = {\rm const},
\end{gather*}
where the constant does not depend on $\omega$, because $\psi_l(R^*_0 - 0)$ is independent of $\omega$.
Hence, $\omega \tan\left(\sqrt{\omega^2 - m_-^2}\, R^*_0 + \phi'\right) = {\rm const}$.
This algebraic equation can be approximately solved for high enough levels:
\begin{gather*}
\omega_n \approx \frac{\pi n}{\left|R^*_0\right|}, \quad {\rm for} \quad n \gg 1.
\end{gather*}
From here it follows that the total number of discrete energy levels can be estimated as $N \approx \frac{m \left|R^*_0\right|}{\pi}$.

\section{Are there discrete levels in the presence of gravitational collapse?}

Following the original works, in \cite{Akhmedov:2015xwa} it was shown that the trajectory of the thin shell in the free fall under its own gravity is $R(t) \approx r_g\left(1 + \frac{R_0 - r_g}{r_g}e^{-t/r_g}\right)$, as $t \to +\infty$ (we assume that $r_g$ is big enough to have an actual collapse without a bounce). Also the relation of the internal ($t_-$) and external ($t$) clock rates is as follows: $t_-  \approx \frac{R_0-r_g}{\nu}\left(1 - e^{-t/r_g}\right)$, as $t \to + \infty$. We substitute these relations into the equation \eqref{bound} and obtain the following boundary conditions for the scalar field during the final stage of the collapse\footnote{One should be careful with these boundary conditions. In fact, both sides of the second equation in (\ref{bound}) can be of the same order, but both of them tend to zero as $t\to +\infty$ during the collapse process \cite{Akhmedov:2015xwa}. At the same time we drop off the first equation because at the final stage of the collapse process the star surface is moved infinitely far away, as measured by the Schwarzschild time $t$, which is used here.}:
	\begin{gather}
	\left(\partial_t -\partial_{r^*}\right)\left.\phi_l(r^*,t)\right|_{r^* = R(t)} \approx 0. \label{newbound}
	\end{gather}
Here we have assumed that the field inside the collapsing star behaves regularly with respect to the internal coordinates, i.e. $\left|\partial_{t_-}\phi\right|$ and $\left|\partial_r \phi\right|$ are finite. At the same time the equations (\ref{eoms}) reduce to
\begin{gather}\label{eqonpsicollapse}
\left[\partial_t^2 - \partial_{r_*}^2\right]\psi_l \approx 0, \quad {\rm when} \quad r \sim R(t), \quad {\rm and } \nonumber \\
\left[\partial_t^2 - \partial_{r_*}^2 - m^2\right]\psi_l \approx 0, \quad {\rm as} \quad r_* \to + \infty.
\end{gather}

Then the asymptotic form of the harmonic functions can be easily found to be:
\begin{gather}\label{asimptot}
\psi_{l,\omega} = \left\{
\begin{matrix}
\alpha e^{-i \omega (r^* + t)} + \beta e^{i \omega (r^* - t)} , r^* \to - \infty,\\
e^{- i \, \omega \, t} \, e^{- \sqrt{m^2 - \omega^2} r^*}, r^* \to +\infty,
\end{matrix}
\right.
\end{gather}
for some constants $\alpha$ and $\beta$.
Here we assume that $\omega < m$, as it should be for the discrete spectrum, if it does exist.
The boundary condition $0 \approx \left(\partial_t -\partial_{r^*}\right) \psi_{l,\omega} = - 2 i \omega \beta e^{i \omega(r^* - t)}$ implies that $\beta = 0$, but leaves $\alpha$ arbitrary. To find it let us calculate the Vronksian of the solution under consideration:
\begin{gather*}
W =  i\left(\bar{\psi}_l\partial_{r^*} \psi_l - \psi_l\partial_{r^*} \bar{\psi}_l\right) = {\rm const}.
\end{gather*}
It has to be constant as a corollary of the equation of motion (\ref{eoms}).
But if one calculates $W$ at $r^* \to \pm \infty$ using \eqref{asimptot}, he gets the following values
\begin{gather}
	W =
\left\{
\begin{matrix}
 - 2 \, \omega \, \left|\alpha\right|^2, r^* \to - \infty\\
 0, r^* \to +\infty
\end{matrix}\right.
\end{gather}
This fixes $\alpha$ to be also zero. Hence, the harmonic functions obeying the boundary conditions (\ref{newbound}) and (\ref{asimptot}) simultaneously, should be equal to zero everywhere. Therefore this observations imply that there are no discrete energy levels in the vicinity of collapsing stars.

\section{Conclusions}

Considerations of this note can be easily extended to other types of boundary conditions at the star surface, to fermions and to other types of the star matter content. Also one can extend them to the rotating case and Oppenheimer--Snyder collapse \cite{Oppenheimer:1939ue}. On general physical grounds it is not hard to see that the conclusion about the existence of the discrete energy levels will be the same. In this paper we addressed the problem
just in the simplest settings, where part of the conclusions can be observed with the use of analytical calculations.

{\bf{Acknowledgements}} We would like to acknowledge discussions with D.Vasiliev, S.Babak and H.Godazgar. AET and FKP would like to thank the AEI, in particular Hermann Nicolai and Stefan Theisen, for their generous hospitality while this project was being completed. The article was prepared within the framework of a subsidy granted to the National
Research University Higher School of Economics by the Government of the Russian Federation for the
implementation of the Global Competitiveness Program. The work of ETA and FKP was partially supported by the grant for the support of the leading scientific schools SSch--1500.2014.2, by their grants from the Dynasty foundation and was done under the partial support of the RFBR grant 15-01-99504. The work of FKP and DAK is done under the partial support of the RFBR grant 16-32-00064-mol-a and by the support from the Ministry of Education and Science of the Russian Federation (Contract No. 02.A03.21.0003 dated of August 28, 2013).

\end{document}